\documentclass[aps,physrev,reprint,notitlepage,nofootinbib,groupedaddress,floatfix]{revtex4-2}

\usepackage[T1]{fontenc}
\usepackage[utf8]{inputenc}
\usepackage{lmodern}
\usepackage{microtype}
\usepackage{amsmath,amssymb,amsthm,mathtools,bm}
\usepackage{graphicx}
\usepackage{booktabs}
\usepackage{xcolor}
\usepackage{hyperref}
\usepackage[nameinlink,capitalise,noabbrev]{cleveref}

\hypersetup{
  colorlinks=true,
  linkcolor=blue!45!black,
  citecolor=blue!45!black,
  urlcolor=blue!45!black,
  pdftitle={Sharp null-energy lower bounds for Page-like entropy excursions in two-dimensional
conformal field theory},
  pdfauthor={Soufiane}
}

\allowdisplaybreaks[3]

\newcommand{\dd}{\,\mathrm d}
\newcommand{\R}{\mathbb R}
\newcommand{\Eplus}{E_+}
\newcommand{\Eminus}{E_-}
\newcommand{\Enet}{E_{\rm net}}
\newcommand{\Calpha}{\mathcal C_\alpha}
\newcommand{\Qgap}{\mathcal Q}
\newcommand{\Rgap}{\mathfrak R_\alpha}
\newcommand{\esssup}{\operatorname*{ess\,sup}}

\begin{document}

\title{Sharp null-energy lower bounds for Page-like entropy excursions in two-dimensional conformal field theory}

\author{Soufiane Karmoudi}
\email{soufiane.elkarmoudi@uit.ac.ma}
\author{Mohammed Daoud}
\affiliation{LPMS, Department of Physics, Faculty of Sciences, Ibn Tofail University, Kenitra, Morocco}

\begin{abstract}
We derive sharp lower bounds on the energy cost of Page-like entanglement entropy excursions in two-dimensional conformal field theory. For states satisfying the primary quantum null energy condition (QNEC), the positive null energy \(E_+\) obeys a sharp lower bound with two branches separated by the threshold \(\Delta S=(c/6)\ln 2\). We also derive a sharp lower bound on the net energy \(E_{\rm net}\), with a coefficient larger by a factor of two than the value quoted by Abdolrahimi and Page \hyperlink{cite.AbdolrahimiPage2015}
{[Phys. Rev. D \textbf{92}, 083005 (2015)]} and, in the QNEC-saturating sector, we obtain a sharp lower bound on the negative energy \(E_-\). The constants are sharp as infima over smooth conformal vacua. We determine the entropy profiles that minimize \(E_+\) for general QNEC states and the one that minimizes \(E_-\) in the saturating sector. Finally, for an excursion containing an infinite sequence of small sub-excursions within a finite duration, we derive a necessary condition for the total positive energy to remain finite, which is stronger than requiring the entropy to have finite total variation.
\end{abstract}

\maketitle

\section{Introduction}
A Page-like entropy profile for the radiation emitted by a black hole is expected in a unitary evaporation and would thus remove at least some of the tension underlying the information paradox \cite{Page1993}.
Moving mirrors in two-dimensional conformal field theory provide a convenient framework to mimic some aspects of black-hole behavior and possibly gain insights into black-hole thermodynamics. Thus, it is relevant to constrain physical processes that would result in Page-like entropy excursions in such a framework before trying to extend them to four-dimensional evaporating black holes. Hence, we asked ourselves the two following questions: 

\begin{quote}
    For a prescribed Page-like excursion of null-deformed entanglement entropy in a 2D CFT during a prescribed affine duration, what are the minimum values of the positive, negative, and net null energies? What are the exact entropy profiles corresponding to those values? 
\end{quote}

Indeed, in 2D CFT, entanglement production and negative energy are linked and quantum null-energy inequalities constrain the physically admissible energy-entropy profiles. For a moving mirror or a conformal vacuum, the conformal anomaly fixes the null-null stress-energy tensor component in terms of a ray-tracing function \cite{FullingDavies1976,DaviesFulling1977} and the same function determines a finite covariantly renormalized entanglement entropy \cite{HolzheyLarsenWilczek1994,BianchiSmerlak2014,BianchiDeLorenzoSmerlak2015}. Eliminating the ray-tracing function gives the Bianchi-Smerlak entropy-flux relation \cite{BianchiSmerlak2014}. Its right-hand side is also the nonlinear entropy combination appearing in the primary two-dimensional quantum null energy condition (QNEC) in the classes in which that condition holds \cite{EckerEtAl2019,KibeEtAl2022,KibeRoy2025}.
Also, a nontrivial conformal-vacuum entropy pulse requires flux of both signs \cite{BianchiSmerlak2014} and explicit Page-like profiles with positive-, zero-, and negative-flux phases were constructed in \cite{AbdolrahimiPage2015}. Moreover, quantum energy inequalities and quantum-interest bounds constrain linear smearings and pulse compensation \cite{FordRoman1999,FewsterTeo2000,FewsterHollands2005} and modular-energy inequalities relate localized positive and negative energy to relative entropy \cite{BlancoEtAl2018}.

To address the two questions raised earlier, we first introduce the conformal entropy-flux relation and the primary QNEC, and then formulate
the corresponding variational problem in the QNEC-saturating sector. This leads to the energy bounds for states satisfying the primary QNEC, their optimal entropy profiles, and the Pareto frontier between the different energy components in the QNEC-saturating sector. We then establish the sharpness of these results over smooth conformal vacua and discuss their moving-mirror and holographic realizations. Finally, we extend the analysis to repeated turnovers and test the bounds with a smooth sinusoidal pulse. The results are applied to a sinusoidal test entropy model in the final section. 

\section{Conformal entropy, null flux, and primary QNEC}

\subsection{Conformal vacua and the Bianchi-Smerlak relation}

Let $(u,v)$ be inertial null coordinates in two-dimensional Minkowski spacetime,
\begin{equation}
\dd s^2=-\dd u\,\dd v.
\end{equation}
A timelike moving mirror is described by
\begin{equation}
v=p(u),\qquad p'(u)>0.
\label{ray-tracing}
\end{equation}
In the conformal-vacuum sector with central charge $c$, the conformal transformation law of the chiral stress tensor gives the Fulling-Davies outgoing null-energy flux \cite{FullingDavies1976,DaviesFulling1977}
\begin{equation}
\langle T_{uu}(u)\rangle=F(u)=-\frac{c}{24\pi}\{p(u),u\},
\label{FD}
\end{equation}
where
\begin{equation}
\{p,u\}=\frac{p'''}{p'}-\frac32\left(\frac{p''}{p'}\right)^2
\end{equation}
is the Schwarzian derivative.

For a chiral interval $[u_0,u]$ where $\epsilon_0$ and $\epsilon$ are UV regulators associated with $u_0$ and $u$ respectively, the inertial-vacuum entropy is \cite{HolzheyLarsenWilczek1994,CalabreseCardy2004}
\begin{equation}
S_{\rm in}(u_0,u)=\frac c{12}\ln\frac{(u-u_0)^2}{\epsilon_0\epsilon},
\end{equation}
while in the conformal vacuum defined by $p$ it is \cite{BianchiSmerlak2014}
\begin{equation}
S_p(u_0,u)=\frac c{12}\ln\frac{[p(u)-p(u_0)]^2}{p'(u_0)p'(u)\epsilon_0\epsilon}.
\end{equation}
The finite entropy is their difference.  Taking $u_0\to-\infty$ in an asymptotically affine region gives
\begin{equation}
S(u)-S_0=-\frac c{12}\ln p'(u).
\label{ptoS}
\end{equation}
Eliminating $p'$ between \eqref{FD} and \eqref{ptoS} yields the Bianchi-Smerlak relation \cite{BianchiSmerlak2014}
\begin{equation}
F_S(u)=\frac{1}{2\pi}
\left[S''(u)+\frac6cS'(u)^2\right].
\label{BS}
\end{equation}
We attach the subscript $S$ because this is the stress profile determined by the entropy in the saturating conformal-vacuum sector.

With
\begin{equation}
K=\frac{c}{12\pi},
\qquad
\psi(u)=\exp\left[\frac6c(S(u)-S_0)\right],
\label{Kpsi}
\end{equation}
one finds
\begin{equation}
F_S(u)=K\frac{\psi''(u)}{\psi(u)}.
\label{linearized flux}
\end{equation}
Equivalently, $\psi$ is a positive zero mode of a one-dimensional Sturm-Liouville operator.

\subsection{Primary QNEC and energy cost}

We assume that the state, region, regulator, and affine null deformation obey the primary two-dimensional QNEC,
\begin{equation}
F(u)
\geq F_S(u) = 
\frac{1}{2\pi}
\left[S''(u)+\frac6cS'(u)^2\right].
\label{primary-QNEC}
\end{equation}
This strengthened form is saturated in the conformal-vacuum sector discussed above and, in holographic CFTs, for states dual to vacuum Bañados geometries \cite{EckerEtAl2019}. Its use outside such classes remains an explicit assumption \cite{KibeRoy2025}.

Setting $[F]_+(u)=max\{F(u),0\}$ and using Jordan decomposition for the energy $\Enet=\Eplus-\Eminus$, write
\begin{equation}
\begin{gathered}
\Enet[F] = \int F\,\dd u,\quad
\Eplus[F] = \int [F]_+\,\dd u,\\
\text{and}\quad
\Eminus[F] = \int [-F]_+\,\dd u.
\end{gathered}
\end{equation}
Then, define
\begin{equation}
F=F_S+\Delta_Q,
\qquad \Delta_Q\geq0,
\qquad
\Qgap=\int\Delta_Q(u)\,\dd u.
\end{equation}
For $0\leq\alpha\leq1$, let
\begin{equation}
\Calpha[F]=\Eplus[F] - \alpha\Eminus[F]. 
\label{Calpha}
\end{equation}
$\Calpha$ is a cost function that measures the positive-energy cost while allowing the negative-energy contribution to offset it with a weight $\alpha$. As $\alpha$ varies from $0$ to $1$, the cost continuously interpolates from $E_+$ to the net energy $\Enet = \Eplus-\Eminus$.
It then follows that (see Appendix~\ref{QNEC})
\begin{equation}
\boxed{
\Calpha[F]
=\Calpha[F_S]+\alpha\Qgap+\Rgap,
}
\end{equation}
with
\begin{equation}
\Rgap=(1-\alpha)\int\bigl([F]_+-[F_S]_+\bigr)\,\dd u.
\end{equation}
$\Rgap\geq0$ as the map $F \mapsto [F]_+$ is nondecreasing, hence $F\geq F_S$ implies $[F]_+\geq [F_S]_+$.

Thus a sharp lower bound on $\Calpha[F_S]$ automatically applies to any state satisfying the primary QNEC. The goal will be to determine such a bound. 

\section{Variational problem in the saturating sector}

\subsection{Complete entropy excursions}

A complete excursion of duration $T$ and height $\Delta S>0$ obeys
\begin{equation}
\begin{aligned}
&S(u)=S_0 && (u\notin[0,T]),\\
&S_0\leq S(u)\leq S_0+\Delta S && (u\in[0,T]).
\end{aligned}
\end{equation}
In terms of \eqref{Kpsi},
\begin{equation}
\begin{aligned}
&\psi(u)=1 && (u\notin[0,T]),\\
&1\leq\psi(u)\leq R && (u\in[0,T]), \qquad R=e^{6\Delta S/c}.
\end{aligned}
\end{equation}
The exact relaxed optimizers have jumps in $\psi'$ and therefore
the flux $F_S$, which depends on $\psi''$ through
\eqref{linearized flux}, contains Dirac $\delta$ contributions.
We thus allow $\psi' \in BV(\mathbb R)$.
Hence, the flux measure is
\begin{equation}
F_S\,\dd u=K\frac{1}{\psi}D^2\psi,
\end{equation}
where $D^2\psi$ is the second distributional derivative of $\psi$ accounting for $\psi'$ jumps. 

Furthermore, it's demonstrated in Appendix~\ref{Enet} that
\begin{equation} 
\int_\R D\left(\frac{\psi'}{\psi}\right)
=\int_\R \frac1\psi D^2\psi-\int_\R\left(\frac{\psi'}{\psi}\right)^2\dd u = 0,
\end{equation}
and as $\psi'=0$ outside the pulse,   
\begin{equation}
\Enet[F_S]
=K\int_\R\frac1\psi D^2\psi
=K\int_0^T\left(\frac{\psi'}{\psi}\right)^2\dd u.
\label{net-action}
\end{equation}

\subsection{Reduction to one monotone rise and fall}

An arbitrary admissible pulse may contain local extrema below $R$. Here we consider only the rising portion of the pulse, connecting $1$ to $R$, with $\psi(u_*)=R$ and $\psi'(u)=0$ for $u\notin(0,u_*)$. Hence, let a Lipschitz profile connect $1$ to $R$, possibly with several upward and downward crossings. For almost every level $y\in(1,R)$ that is crossed upward one or more times, let $u_i(y)$ denote the upward-crossing points, so that 
\begin{equation}
    \psi(u_i(y))=y, \qquad \psi'(u_i(y))>0,
\end{equation} 
and define their crossing speeds by 
\begin{equation}
    v_i(y)=\psi'(u_i(y))>0. 
\end{equation}
From $\dd y=\psi'(u)\dd u=v_i(y)\dd u$, the total time spent crossing upward the interval $\dd y$ is \begin{equation} 
\dd T_\uparrow = \frac{\dd y}{v_1} +\frac{\dd y}{v_2}+\cdots=\sum_i\frac{\dd y}{v_i}, \end{equation} 
hence the total time during which $\psi$ is increasing is \begin{equation} T_\uparrow=\int_1^R\sum_i\frac{\dd y}{v_i(y)}. \end{equation} 
The energy of this non-monotone profile can be decomposed as 
\begin{equation} E_{\rm net}^{\rm nm} = K\int_{\{\psi'>0\}} \left(\frac{\psi'}{\psi}\right)^2 \dd u + K\int_{\{\psi'< 0\}} \left(\frac{\psi'}{\psi}\right)^2 \dd u,
\end{equation} 
changing $\psi$ into $y$ gives 
\begin{equation} 
K\int_{\{\psi'>0\}}\left(\frac{\psi'}{\psi}\right)^2\dd u =K\int_1^R\frac{\sum_iv_i(y)}{y^2}\,\dd y \leq E_{\rm net}^{\rm nm}. \end{equation} 
Now we want to compare this profile to a monotone increasing one that takes the same time $T_\uparrow$ to connect 1 to R. We therefore define a unique crossing speed $p(y)$ (that should not be confused with the ray-tracing function $p(u)$ in \eqref{ray-tracing}) for the interval $\dd y$ that yields the same crossing time $\dd T_\uparrow$: 
\begin{equation} \sum_i\frac{\dd y}{v_i} = \frac{\dd y}{p(y)} \iff p(y)=\left(\sum_i\frac1{v_i(y)}\right)^{-1}, \end{equation} 
so that 
\begin{equation} \int_1^R\frac{\dd y}{p(y)}=T_\uparrow. 
\end{equation} 
For positive $v_i(y)$, 
\begin{equation} \left(\sum_i \frac{1}{v_i(y)}\right)^{-1} \leq \sum_i v_i(y), \end{equation}  
such as the monotone profile satisfies
\begin{equation} E_{\rm net}^{\rm m} = K\int_1^R \frac{p(y)}{y^2}\,\dd y \leq K\int_1^R \frac{\sum_i v_i(y)}{y^2}\,\dd y \leq E_{\rm net}^{\rm nm}. 
\end{equation}

Moreover, in the non-monotone regime, let
\begin{equation}
    M = \operatorname*{ess\,sup}_{\{\psi'>0\}}\psi'(u), \quad v_i(y)\leq M, 
\end{equation}
consequently 
\begin{equation}
    \sum_i\frac{1}{v_i(y)}\geq\frac{1}{M} \iff p(y)\leq M. 
\end{equation}
The effective speed $p(y)$ therefore defines a monotone profile connecting $1$ to $R$ in time $T_\uparrow$, which is no longer than
the total duration of the non-monotone profile, with no larger maximal speed
and no larger energy. The same conclusions apply to the right side connecting $R$ to $1$ after reversing its time parametrization.

Thus, a minimizing complete pulse consists of one monotone rise and one monotone fall. Constant phases at $\psi=1$ or $\psi=R$ may still occur. If the durations of the rise and fall are $\tau_L$ and $\tau_R$ respectively, then $\tau_L+\tau_R\leq T$.
\subsection{One-sided optimization}

Consider again one monotone segment of duration $\tau$ joining $\psi=1$ to $\psi=R$. Use $\psi$ as the independent variable and set
\begin{equation}
p(\psi)=\frac{\dd\psi}{\dd u}>0,
\qquad
M=\esssup_{1<\psi<R}p(\psi).
\end{equation}
The duration is
\begin{equation}
\tau=\int_1^R\frac{\dd\psi}{p(\psi)},
\end{equation}
and \eqref{net-action} becomes
\begin{equation}
\Enet[F_S]=K\int_1^R\frac{p(\psi)}{\psi^2}\,\dd\psi.
\end{equation}
Because $\psi'$ vanishes before and after the pulse, the downward variation of $p$ is at least $M$, so that the measure 
\begin{equation}
    (D^2\psi)^-[0,\tau] \geq M
\end{equation}
using Jordan decomposition $D^2\psi=(D^2\psi)^+ - (D^2\psi)^-$. 
Then, as $\psi\leq R$
\begin{equation}
    \Eminus[F_S] = K\int_0^{\tau}\frac{1}{\psi}(D^2\psi)^-\geq K\int_0^{\tau}\frac{1}{R}(D^2\psi)^-,
\end{equation}
and as $\int_0^{\tau}(D^2\psi)^-=(D^2\psi)^-[0,\tau]$, one finally gets 
\begin{equation}
\Eminus[F_S]\geq\frac{KM}{R}.
\label{Eminus1}
\end{equation}
Equality is achieved in the relaxed class by concentrating the complete downward jump at $\psi=R$.

From \eqref{Calpha}, $\Calpha=\Enet+(1-\alpha)\Eminus$, therefore
\begin{equation}
\Calpha[F_S]\geq K\left[
\int_1^R\frac{p(\psi)}{\psi^2}\,\dd\psi
+(1-\alpha)\frac{M}{R}
\right].
\label{one side cost}
\end{equation}
At fixed $(M,\tau)$, we have here an optimization problem that consists of minimizing $\int_1^R(p(\psi)/\psi^2)\dd\psi$ while respecting the constraints $\int_1^R\dd\psi/p(\psi)=\tau$ and $p(\psi)\leq M$. The corresponding KKT Lagrangian is
\begin{equation}
\begin{aligned}
L[p,\lambda,\mu]
={}&
\int_1^R \frac{p(\psi)}{\psi^2}\,\dd\psi
+
\lambda\left(
\int_1^R\frac{\dd\psi}{p(\psi)}-\tau
\right)
\\
&+
\int_1^R
\mu(\psi)\bigl[p(\psi)-M\bigr]\,\dd\psi .
\end{aligned}
\end{equation}
Stationarity with respect to $p$ gives 
\begin{equation}
    \frac{\partial \mathcal{L}}{\partial p} = \frac{1}{\psi^2}-\frac{\lambda}{p^2}+\mu=0,
\end{equation}
while the KKT conditions impose
\begin{equation}
    \mu\geq0,\qquad M-p\geq0, \qquad \mu(M-p)=0.
\end{equation}
This results into two possibilities. The first corresponds to $\mu=0$, the inequality constraint is inactive and hence $p(\psi)<M$, in which
case $p(\psi)=a\psi$, where $a=\sqrt{\lambda}$. The second possibility corresponds to $\mu>0$, thus the constraint is active and $p(\psi)=M$. Hence,
\begin{equation}
\boxed{p_*(\psi)=\min\{a\psi,M\}.}
\end{equation}
Let
\begin{equation}
z=\frac{M}{a}\in[1,R].
\end{equation}
Then
\begin{equation}
p_*(\psi)=
\begin{cases}
a\psi,&1\leq\psi\leq z,\\
M,&z\leq\psi\leq R.
\end{cases}
\end{equation}
It follows then that 
\begin{equation}
\tau
=
\int_1^z \frac{\dd \psi}{a\psi}
+
\int_z^R \frac{\dd \psi}{M} = \frac{R + z\ln z - z}{M}
\label{tau}
\end{equation}
And \eqref{one side cost} becomes 
\begin{equation}
\Calpha[F_S]\geq\frac{K}{\tau}f_{\alpha,R}(z),
\label{fbound}
\end{equation}
where
\begin{equation}
f_{\alpha,R}(z)=
\left(R+z\ln z-z\right)\left(\frac{1+\ln z}{z}-\frac{\alpha}{R}\right).
\end{equation}

\subsection{Exact minimization}

Differentiating,
\begin{equation}
f'_{\alpha,R}(z)
=-\frac{\ln z}{Rz^2}
\left(R^2-2Rz+\alpha z^2\right).
\label{fprime}
\end{equation}
A first minimum is at $z=1$. Then, if $z>1$, \eqref{fprime} admits two roots, $z_{\pm} = R/(1\pm q)$ where $q=\sqrt{1-\alpha}\in[0,1]$. However, $z_-\geq R$, thus only remains $z_+$ corresponding to $R\geq 1+q$. 
Let $g_\alpha(R)={\rm min} f_{\alpha,R}(z)$, the solutions write 
\begin{equation}
g_\alpha(R)=
\begin{cases}
\dfrac{(R-1)(R-\alpha)}{R},
&1\leq R\leq1+q,\\[2mm]
\left[q+\ln\!\left(\dfrac{R}{1+q}\right)\right]^2,
&R\geq1+q,
\end{cases}
\end{equation}

Hence, a monotone entropy excursion of affine duration $\tau$ and amplitude $\Delta S$, with endpoint-stationary continuation, under the primary QNEC, satisfies 
\begin{equation}
\Calpha[F]\geq\frac{K}{\tau}g_\alpha(e^{6\Delta S/c})+\alpha\Qgap+\Rgap,
\qquad 0\leq\alpha\leq1.
\label{one side excursion}
\end{equation}
The absolute lower bound is attained in the relaxed saturating class and is the infimum over smooth conformal vacua as shown in \cref{realizations}.

\section{Complete pulses: energy bounds and optimal entropy profiles}

\subsection{Energy bounds}

Now we study the complete pulse, i.e. the rising, falling and potentially constant portions together. Let $\tau_L$ and $\tau_R$ the durations of the rise and fall, respectively, while $T$ is the duration of the complete pulse. From $\tau_L+\tau_R\leq T$, one has
\begin{equation}
\frac1{\tau_L}+\frac1{\tau_R}
\geq\frac4{\tau_L+\tau_R}
\geq\frac4T.
\end{equation}
Hence, the time allocation minimizing the energetic cost $C_{\alpha}$ is $\tau_L=\tau_R=T/2$, allowing for no constant phases. 

Adding the energetic costs of the rise and fall, the total cost of a complete entropy excursion of height $\Delta S$ and affine duration $T$ satisfying primary QNEC obeys 
\begin{equation}
\Calpha[F]
\geq
\frac{c}{3\pi T}
g_\alpha\!\left(e^{6\Delta S/c}\right)+\alpha\Qgap+\Rgap.
\label{costbound}
\end{equation}

\begin{figure}
\includegraphics[width=\columnwidth]{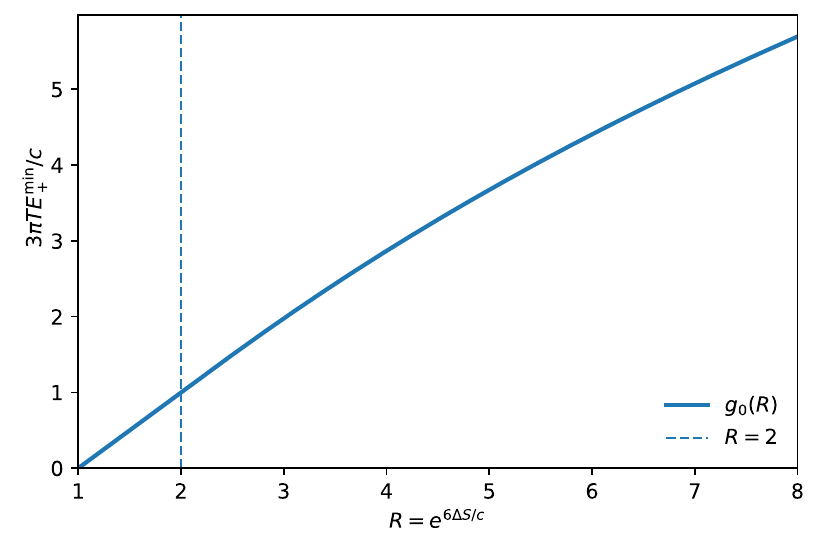}
\caption{Dimensionless sharp positive-energy cost $g_0(R)=3\pi T E_+^{\min}/c$.  The optimizer changes active set at $R=2$, or $\Delta S=(c/6)\ln2$.}
\label{g0}
\end{figure}

Now we set $\Qgap=\Rgap=0$ to obtain the absolute energy bounds for all states satisfying the primary QNEC. At $\alpha=0$, $q=1$, the threshold is $R=2$ or $\Delta S_c=(c/6)\ln2$ and $\Calpha$ simplifies into $\Eplus$, so that
\begin{equation}
\boxed{
\Eplus[F]
\geq
\frac{c}{3\pi T}
\begin{cases}
\exp(6\Delta S/c)-1,
&\Delta S\leq(c/6)\ln2,\\[1mm]
\left(1+6\Delta S/c-\ln2\right)^2,
&\Delta S\geq(c/6)\ln2,
\end{cases}}
\end{equation}
while for $\alpha=1$, $\Calpha$ simplifies into $\Enet$,
\begin{equation}
\boxed{
\Enet[F]\geq\frac{12\Delta S^2}{\pi cT}}.
\end{equation}
First, it should be noted that the absolute minimum of $\Eplus$ can be attained by states that do not saturate the QNEC. Indeed, setting $\alpha=0$ in \eqref{costbound} makes the term $\alpha\Qgap$ vanish, so that only $\Rgap$ must vanish. Also, the lower bound on $\Eplus$ has two branches, with a threshold at $\Delta S=(c/6)\ln 2$. This is highlighted in \cref{g0}.
Moreover, the absolute minimum of $\Enet$ is larger by a factor of two than the value quoted in \cite{AbdolrahimiPage2015}.

Regarding $\Eminus$, we cannot infer from this work a universal lower bound for the states satisfying the primary QNEC. The bounds on $\Eplus$ and $\Enet$ can be lifted from the QNEC-saturating sector because $F(u)\geq F_S(u)$ which implies $[F(u)]_+\geq [F_S(u)]_+$,
and, after integration, $\Eplus[F]\geq \Eplus[F_S]$ and $\Enet[F]\geq \Enet[F_S]$. For the negative-energy part, however, the inequality is reversed:
\begin{equation}
    [F(u)]_-\leq [F_S(u)]_-.
\end{equation}
Hence, a lower bound on $\Eminus$ established in the saturating sector cannot in general be extended to the full primary-QNEC class using the same argument.
Setting $\tau=T/2$ in \eqref{tau}, one gets 
\begin{equation}
    M = \frac{2(R + z\ln z - z)}{T}.
\end{equation}
For a complete pulse, $\eqref{Eminus1}$ therefore becomes
\begin{equation}
    \Eminus[F_S]\geq \frac{c}{3\pi T R}
\left(R+z\ln z-z\right).
\label{Eminus2}
\end{equation}
The minimum is reached at $z=1$
\begin{equation}
\boxed{
    \Eminus[F_S] \geq \frac{c}{3\pi T} \left(1-e^{-6\Delta S/c}\right)}.
\end{equation}
This is therefore the minimum negative-energy cost of a complete entropy pulse of height $\Delta S$ and affine duration $T$ for a state saturating the primary QNEC.

\subsection{Optimal entropy profiles}

As to the entropy evolution throughout the pulse, since $\psi'(u)=p_*(\psi)=\min\{a\psi,M\}$, with the switch from $a\psi$ to $M$ occurring at $\psi=z$, one has (for $\bar u=\min\{u,T-u\}$)
\begin{equation}
\psi(u)=
\begin{cases}
e^{a\bar u},
& 0\leq \bar u\leq u_z,\\[1mm]
z+M(\bar u-u_z),
& u_z\leq \bar u\leq T/2,
\end{cases}
\qquad
u_z=\frac{\ln z}{a},
\end{equation}
while the entropy obeys
\begin{equation}
S(u)=S_0+
\begin{cases}
\dfrac{ca}{6}\,\bar u,
& 0\leq \bar u\leq u_z,\\[2mm]
\dfrac{c}{6}\ln\!\left[z+M(\bar u-u_z)\right],
& u_z\leq \bar u\leq T/2.
\end{cases}
\label{entropy expressions}
\end{equation}

\begin{figure*}
    \centering
    \includegraphics[width=\textwidth]{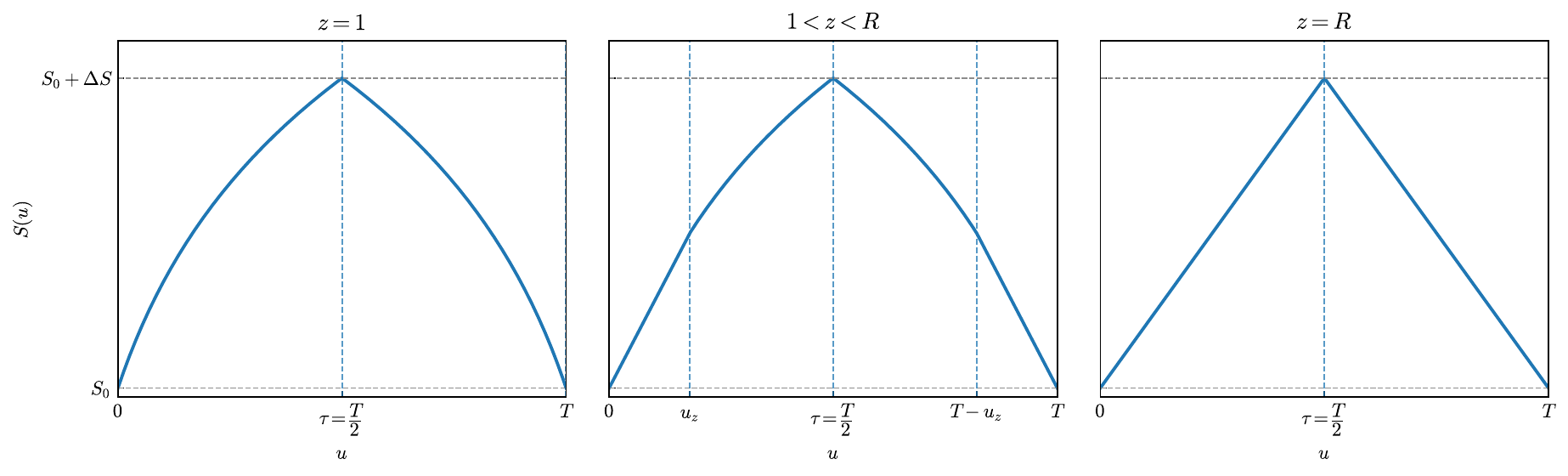}
    \caption{Optimal entropy profiles for the three variational regimes: the logarithmic profile for $z=1$, the mixed linear-logarithmic profile for $1<z<R$, and the linear profile for $z=R$.}
    \label{entropy profiles}
\end{figure*}

As represented in \cref{entropy profiles}, minimizing the cost $\Calpha$ yields three possible entropy profiles: a full logarithmic profile corresponding to the second line of \eqref{entropy expressions} with $u_z=0$, a mixed linear-logarithmic profile corresponding to both lines of \eqref{entropy expressions} with $0<u_z<T/2$, or a full linear profile corresponding to the first line of \eqref{entropy expressions} with $u_z=T/2$. 

Minimizing $\Eminus$ favors the logarithmic profile, whereas $\Enet$ is optimized by the linear one. On the other hand, minimizing $\Eplus$ selects the logarithmic profile for $\Delta S\leq(c/6)\ln2$ and the mixed one for $\Delta S\geq(c/6)\ln2$. Hence, below the threshold, the same logarithmic configuration is optimal for both $\Eminus$ and $\Eplus$, but not for $\Enet$, since $\Enet=\Eplus-\Eminus$ and a larger negative-energy contribution can therefore lower the net energy.

According to the Bianchi-Smerlak relation \eqref{BS}, the linear phases of the entropy, whether increasing or decreasing, correspond to a strictly positive flux, while the logarithmic phases correspond to a vanishing flux. In both cases, the changes of slope at the two endpoints produce positive impulses, and the reversal of the slope at $u=T/2$ produces a negative impulse.

\section{The saturating-sector Pareto frontier}

To better capture the trade-offs between the different energy components, we focus here on the primary QNEC-saturated sector. From \eqref{fbound} and \eqref{Eminus2}, let
\begin{equation}
\begin{aligned}
E_{\rm net}^S(z)
&=
\frac{c}{3\pi T}
\left(R+z\ln z-z\right)
\left(
\frac{1+\ln z}{z}-\frac{1}{R}
\right),\\[1mm]
E_{+}^S(z)
&=
\frac{c}{3\pi T}
\left(R+z\ln z-z\right)
\frac{1+\ln z}{z},\\[1mm]
E_{-}^S(z)
&=
\frac{c}{3\pi T}
\frac{R+z\ln z-z}{R}.
\end{aligned}
\end{equation}

\begin{figure}
\includegraphics[width=\columnwidth]{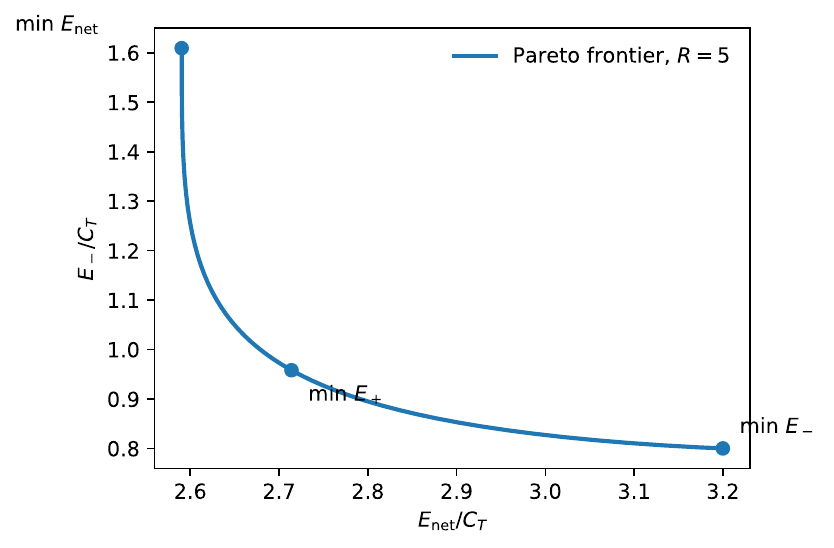}
\caption{QNEC-saturating Pareto frontier for $R=5$, with $C_T=c/(3\pi T)$.}
\label{pareto}
\end{figure}

As mentioned earlier, for $R\leq2$, $z=1$ minimizes both $\Eplus^S$ and $\Eminus^S$. For $R>2$, no common minimizer exists and approaching one of the three energies to its minimum does the opposite for the two others. This is highlighted in \cref{pareto}, where the Pareto frontier in the $(\Enet^S,\Eminus^S)$ plane is plotted for $R=5$.

\section{Sharpness and physical realizations}
\label{realizations}

\subsection{Smooth approximation of the relaxed optimizers}

So far, we have considered relaxed profiles for $S$ or $\psi$ with jumps in their derivatives. Here we prove that the relaxed values remain sharp infima over smooth conformal vacua. 

Let $\psi_*$ be a relaxed one-segment optimizer and consider the monotone rise from $\psi(0)=1$ to $\psi(\tau)=R$ such that $v_*=\psi_*'\geq0$. Then, compress the profile $v_*$ into $[2\varepsilon,\tau-2\varepsilon]$, preserving its total rise, and denote the compressed profile by $\bar v_\varepsilon$.  Let $\rho_\varepsilon$ be a nonnegative smooth mollifier supported in $[-\varepsilon,\varepsilon]$ and define
\begin{equation}
v_\varepsilon=\rho_\varepsilon*\bar v_\varepsilon,
\qquad
\psi_\varepsilon(u)=1+\int_{-\infty}^uv_\varepsilon(s)\,\dd s.
\end{equation}
Then $v_\varepsilon\geq0$, $\psi_\varepsilon$ is smooth, $\psi_\varepsilon=1$ before the segment and $\psi_\varepsilon=R$ after it. As $\varepsilon \to 0$, one has the convergence properties
\begin{equation}
\begin{gathered}
\|\psi_\varepsilon-\psi_*\|_\infty \to 0, \\[1mm]
D^2\psi_\varepsilon \rightharpoonup^* D^2\psi_*
\rlap{$\qquad (\varepsilon\downarrow 0)$}, \\[1mm]
\|D^2\psi_\varepsilon\|_{\rm TV}
\to
\|D^2\psi_*\|_{\rm TV}.
\end{gathered}
\end{equation}
The last two properties are equivalent to strict convergence of the second derivative measures, which implies convergence of the Jordan masses $(D^2\psi_\varepsilon)^\pm$. Together with the first property, they allow us to write
\begin{equation}
    K\int \frac{1}{\psi_\varepsilon}\,(D^2\psi_\varepsilon)^\pm \longrightarrow K\int \frac{1}{\psi_*}\, (D^2\psi_*)^\pm \qquad (\varepsilon\downarrow 0),
\end{equation}
hence
\begin{equation}
    E_\pm[\psi_\varepsilon]
\longrightarrow
E_\pm[\psi_*]
\qquad
(\varepsilon\downarrow 0).
\end{equation}
For a finite profile composed of several segments, the same smoothing procedure can be applied independently to each segment, with descending segments treated analogously by time reversal and with non-overlapping smoothing regions.

For every smooth positive approximant, \eqref{ptoS} gives
\begin{equation}
p_\varepsilon'(u)=\psi_\varepsilon(u)^{-2}>0.
\end{equation}
It is straightforward that 
\begin{equation}
    S-S_0=\frac{c}{6}\ln\psi_\varepsilon,
\qquad
F=K\frac{\psi_\varepsilon''}{\psi_\varepsilon}=F_S.
\end{equation}
Hence, each smooth approximant is realized by a QNEC-saturating conformal vacuum.

Therefore, the minimum of $\Calpha$ in the relaxed class is equal to its infimum in the smooth class, realized here by smooth conformal vacua.

\subsection{Moving mirrors and Bañados states}

Replacing $p'$ by $\psi^{-2}$ yields the moving mirror velocity 
\begin{equation}
v_m=\frac{p'-1}{p'+1}
=-\tanh(\ln\psi),
\end{equation}
satisfying $|v_m|<1$ and hence corresponding indeed to a timelike trajectory. The rapidity and signed proper acceleration are
\begin{equation}
\eta_m=-\ln\psi,
\qquad
a_m=-\psi'.
\end{equation}
Thus, the endpoint impulses in the relaxed class, arising from jumps in $\psi'$, can be seen as limits of smooth and increasingly sharp changes in acceleration in the smooth class.  

In holographic CFT$_2$, all states dual to vacuum Bañados geometries saturate primary QNEC \cite{EckerEtAl2019}.  In particular, the flux is related to the Bañados function $L_+$ according to $F = cL_+/12\pi$, such that
\begin{equation}
L_+(u)=\frac{\psi''(u)}{\psi(u)}.
\end{equation}
This provides a local holographic saturation framework.

\section{Repeated turnovers and finite positive-energy condition}

Here, we extend the study to a non-monotone entropy excursion with multiple local extrema noted $S_0,S_1,\ldots,S_N$. Each segment linking $S_{i-1}$ to $S_i$ is strictly monotone and lasts a duration $\tau_i$. Constant phases are possible and the overall excursion duration is $T$. Define 
\begin{equation}
\begin{gathered}
\Delta S_i = |S_i-S_{i-1}|,\quad
R_i=e^{6\Delta S_i/c},\\
\text{and}\quad
w_i=\sqrt{g_\alpha(R_i)}.
\end{gathered}    
\end{equation}

Applying \eqref{one side excursion} yields the total energetic cost (see Appendix~\ref{rearrangement})  
\begin{equation}
\Calpha[F]
\geq
\frac{K}{T}\left(\sum_{i=1}^N w_i\right)^2
+\alpha\Qgap+\Rgap.  
\end{equation} 
The optimizer requires the absence of constant phases and the time allocation for each segment to be (Appendix~\ref{rearrangement})
\begin{equation}
\tau_i^{\rm opt}=T\frac{w_i}{\sum_kw_k}.
\end{equation}
If the overall excursion contains $n$ segments of the same height $\Delta S$, then the minimum energy cost (hence in the saturated sector) is related to that of a single segment by
\begin{equation}
\Calpha^{(n),min}=n^2\Calpha^{(1),min}, 
\end{equation}
thus
\begin{equation}
\Eplus^{(n),min}=n^2\Eplus^{(1),min}, \quad \Enet^{(n),min}=n^2\Enet^{(1),min}.      
\end{equation}

Moreover, for small segments, $x=6\Delta S/c \to 0$, such that, for $\alpha<1$, 
\begin{equation}
w(e^x)=\sqrt{1-\alpha}\,\sqrt{x}+O(x^{3/2}). 
\end{equation}
Hence, for an excursion made of an infinity of small segments within a finite total time $T$, even if the total entropy variation is finite, that is 
\begin{equation}
    \sum_i x_i <\infty,
\end{equation}
the total minimum cost $\Calpha^{\rm min}$, and hence $\Eplus^{\rm min}$, can still diverge as
\begin{equation}
    \Calpha^{\rm min} \propto \frac{1}{T} \left(\sum_i\sqrt{x_i}\right)^2.
\end{equation}
For example, if $x_i=1/i^2$, then 
\begin{equation}
    \sum_{i=1}^{\infty}\frac{1}{i^2} = \frac{\pi^2}{6}, \qquad \sum_{i=1}^{\infty}\frac{1}{i} = \infty.    
\end{equation}

Although an infinite number of segments is physically idealized, this argument shows that a highly oscillatory entropy profile with small turnovers requires a large amount of positive energy and, especially, requires the condition
\begin{equation}
\sum_i\sqrt{x_i}<\infty,
\end{equation}
which is stronger than the finite entropy variation condition $\sum_ix_i<\infty$.

\section{Test with a sinusoidal pulse}

\begin{figure}
\includegraphics[width=\columnwidth]{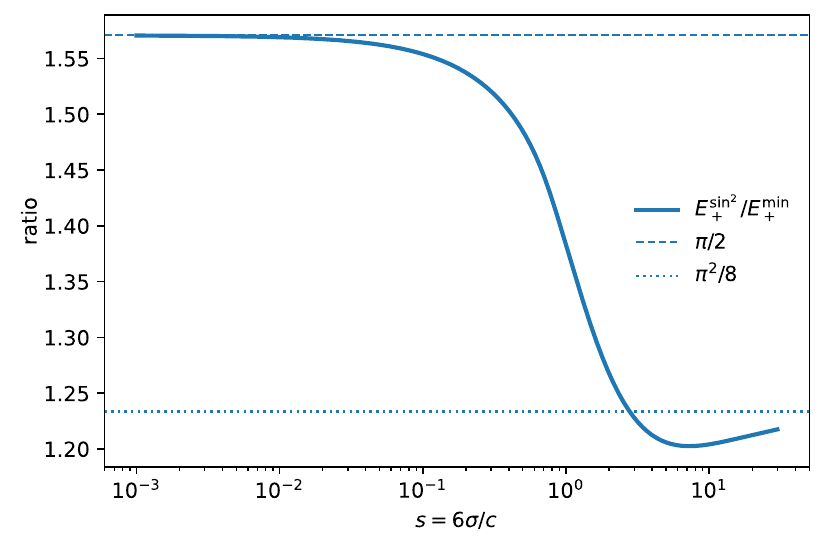}
\caption{Positive-energy ratio of the smooth conformal pulse \eqref{sinusoidalS}.  The dashed and dotted lines are the analytic limits $\pi/2$ and $\pi^2/8$.}
\label{fig:sinusoidal}
\end{figure}

To test the bound on a smooth nonoptimal pulse, consider
\begin{equation}
S(u)-S_0=\sigma\sin^2\left(\frac{\pi u}{T}\right),
\qquad 0\leq u\leq T,
\label{sinusoidalS}
\end{equation}
with constant continuation outside.  Let
\begin{equation}
s=\frac{6\sigma}{c},
\qquad x=\frac{\pi u}{T}.
\end{equation}
The saturating flux is
\begin{equation}
F_S(u)=\frac{\hbar c\pi s}{6T^2}
\left[\cos(2x)+\frac{s}{2}\sin^2(2x)\right].
\label{sinusoidal flux}
\end{equation}
Let
\begin{equation}
r=\frac{s}{\sqrt{1+s^2}+1},
\qquad
\varphi=\arccos r,
\end{equation}
the positive region of $F_S$ is given by
\begin{equation}
x\in
\left[0,\frac{\pi-\varphi}{2}\right]
\cup
\left[\frac{\pi+\varphi}{2},\pi\right].
\end{equation}
Using the symmetry of the flux,
\begin{equation}
\begin{aligned}
E_+^{\sin^2}
&=
\frac{\hbar c s}{3T}
\int_0^{(\pi-\varphi)/2}
\left[
\cos(2x)+\frac{s}{2}\sin^2(2x)
\right]\dd x
\\
&=
\frac{\hbar c s}{6T}
\left[
\sin\varphi
+\frac{s}{8}
\left(
2\pi-2\varphi+\sin2\varphi
\right)
\right].
\end{aligned}
\end{equation}
From \eqref{costbound}, the sharp bound on the positive energy is
\begin{equation}
    E_+^{\min} = \frac{c}{3\pi T}g_0(e^s). 
\end{equation}
Therefore, the ratio to the sharp bound tends to
\begin{equation}
\frac{E_+^{\sin^2}}{E_+^{\min}}\longrightarrow\frac\pi2
\quad(s\to0),
\qquad
\frac{E_+^{\sin^2}}{E_+^{\min}}\longrightarrow\frac{\pi^2}{8}
\quad(s\to\infty).
\end{equation}
Figure~\ref{fig:sinusoidal} shows that the sinusoidal pulse remains within an order-one factor of the sharp bound over the full range of $s$.

As to the net energy, integrating the flux \eqref{sinusoidal flux} gives
\begin{equation}
E_{\rm net}^{\sin^2}
=
\frac{\pi c s^2}{24T},
\end{equation}
while the sharp bound is
\begin{equation}
E_{\rm net}^{\min}
=
\frac{c s^2}{3\pi T}.
\end{equation}
Hence
\begin{equation}
\frac{E_{\rm net}^{\sin^2}}{E_{\rm net}^{\min}}
=
\frac{\pi^2}{8},
\end{equation}
which is independent of $s$, as expected, since the sharp bound on $\Enet$, unlike that on $\Eplus$, does not change form with the excursion height.

\section{Discussion}

Under $\widetilde u=au+b$, one has $\widetilde T=aT$ and $\widetilde E=E/a$, so $TE$ is invariant. However, a general conformal reparametrization adds a Schwarzian term and can change the positive and negative parts of the flux separately.

The bounds derived in this work are complementary to existing energy inequalities \cite{FewsterHollands2005, BlancoEtAl2018, FlissRolph2025}, since they are conditioned on a prescribed entropy excursion and affine duration.

Several extensions are natural.  A separate analysis is needed for gravitationally anomalous CFTs, excess energy over thermal backgrounds, and global Bañados orbit constraints.

\appendix

\section{Primary QNEC and energy cost}
\label{QNEC}
\subsection{Primary-QNEC lift of the cost functional}

From \eqref{Calpha} and $E_{\rm net}=E_+-E_-$, one has
\begin{equation}
\begin{aligned}
\Calpha[F]
&=E_+[F]-\alpha E_-[F] \\
&=\alpha E_{\rm net}[F]+(1-\alpha)E_+[F] \\
&=\alpha\int F\,\dd u
 +(1-\alpha)\int[F]_+\,\dd u.
\end{aligned}
\end{equation}
Since
\begin{equation}
F=F_S+\Delta_Q,
\qquad
Q=\int\Delta_Q\,\dd u,
\end{equation}
it follows that
\begin{align}
\Calpha[F]
={}&
\alpha\int F_S\,\dd u
+(1-\alpha)\int[F_S]_+\,\dd u
+\alpha Q
\nonumber\\
&+(1-\alpha)
\int\left([F]_+-[F_S]_+\right)\dd u.
\end{align}
The first two terms give $\Calpha[F_S]$. Therefore,
\begin{equation}
\Calpha[F]
=
\Calpha[F_S]
+\alpha Q
+\mathfrak R_\alpha,
\end{equation}
where
\begin{equation}
\mathfrak R_\alpha=(1-\alpha)\int\left([F]_+-[F_S]_+\right)\dd u\geq0.
\end{equation}
The last inequality follows from $F\geq F_S$ and the monotonicity of
$F\mapsto[F]_+$.

\section{Net energy integral expression}
\label{Enet}
Since $\psi'\in BV(\mathbb R)$ and $\psi\geq1$, the function
$1/\psi$ is Lipschitz. The BV product rule gives
\begin{equation}
D\left(\frac{\psi'}{\psi}\right)
=
\frac{1}{\psi}D^2\psi
+
\psi' D\left(\frac{1}{\psi}\right).
\end{equation}
As $\psi$ is continuous,
\begin{equation}
D\left(\frac{1}{\psi}\right)
=
-\frac{\psi'}{\psi^2}\,\dd u,
\end{equation}
and hence
\begin{equation}
D\left(\frac{\psi'}{\psi}\right)
=
\frac{1}{\psi}D^2\psi
-
\left(\frac{\psi'}{\psi}\right)^2\dd u.
\end{equation}

Because $\psi=1$ outside the pulse, $\psi'=0$ there, and therefore
$\psi'/\psi$ vanishes at both ends of $\mathbb R$, such that
\begin{equation}
\int_{\mathbb R}
D\left(\frac{\psi'}{\psi}\right)
=0.
\end{equation}

\section{Monotone rearrangement and global reduction}
\label{rearrangement}

From \eqref{one side excursion}, on has first
\begin{equation}
\Calpha[F]
\geq
K\sum_{i=1}^N\frac{w(R_i)^2}{\tau_i}
+\alpha\Qgap+ \Rgap,
\end{equation}
where $\mathcal Q$ and $\mathfrak R_\alpha\geq0$ denote the gaps integrated over the complete interval. Let
\begin{equation}
T_a=\sum_{i=1}^N\tau_i\leq T
\end{equation}
be the total duration of the monotone segments, hence not including the constant phases. By the Cauchy-Schwarz inequality,
\begin{equation}
\begin{aligned}
\left(\sum_{i=1}^N w(R_i)\right)^2
&=
\left(
\sum_{i=1}^N
\frac{w(R_i)}{\sqrt{\tau_i}}\sqrt{\tau_i}
\right)^2
\\
&\leq
\left(
\sum_{i=1}^N\frac{w(R_i)^2}{\tau_i}
\right)
\left(
\sum_{i=1}^N\tau_i
\right)
\\
&=
T_a
\sum_{i=1}^N\frac{w(R_i)^2}{\tau_i}.
\end{aligned}
\end{equation}
Therefore,
\begin{equation}
\sum_{i=1}^N\frac{w(R_i)^2}{\tau_i}
\geq
\frac{\left(\sum_{i=1}^N w(R_i)\right)^2}{T_a}
\geq
\frac{\left(\sum_{i=1}^N w(R_i)\right)^2}{T},
\end{equation}
such that
\begin{equation}
\Calpha[F]
\geq
\frac{K}{T}
\left(
\sum_{i=1}^N w(R_i)
\right)^2 +\alpha\Qgap+ \Rgap.
\end{equation}

Equality in
\begin{equation}
\frac{\left(\sum_{i=1}^N w(R_i)\right)^2}{T_a}
\geq
\frac{\left(\sum_{i=1}^N w(R_i)\right)^2}{T}
\end{equation}
requires
\begin{equation}
T_a=T,
\end{equation}
so no affine time is spent in constant phases.
Then, equality in the Cauchy-Schwarz inequality requires the two sequences
\[
\left\{\frac{w(R_i)}{\sqrt{\tau_i}}\right\}_{i=1}^N
\qquad\text{and}\qquad
\left\{\sqrt{\tau_i}\right\}_{i=1}^N
\]
to be proportional. Hence there exists a constant $\lambda>0$ such that
\begin{equation}
\frac{w(R_i)}{\sqrt{\tau_i}}
=
\lambda\sqrt{\tau_i},
\end{equation}
or equivalently,
\begin{equation}
\tau_i=\frac{w(R_i)}{\lambda}.
\end{equation}
Using $\sum_{i=1}^N\tau_i=T$ then gives
\begin{equation}
\lambda
=
\frac{\sum_{k=1}^N w(R_k)}{T},
\end{equation}
and therefore
\begin{equation}
\tau_i^{\rm opt}
=
T\,
\frac{w(R_i)}
{\sum_{k=1}^N w(R_k)}.
\end{equation}

\bibliography{References}

@article{Page1993,
  author  = {Page, Don N.},
  title   = {Information in Black Hole Radiation},
  journal = {Physical Review Letters},
  volume  = {71},
  pages   = {3743--3746},
  year    = {1993},
  doi     = {10.1103/PhysRevLett.71.3743}
}

@article{FullingDavies1976,
  author  = {Fulling, Stephen A. and Davies, Paul C. W.},
  title   = {Radiation from a Moving Mirror in Two Dimensional Space-Time: Conformal Anomaly},
  journal = {Proceedings of the Royal Society of London A},
  volume  = {348},
  pages   = {393--414},
  year    = {1976},
  doi     = {10.1098/rspa.1976.0045}
}

@article{DaviesFulling1977,
  author  = {Davies, Paul C. W. and Fulling, Stephen A.},
  title   = {Radiation from Moving Mirrors and from Black Holes},
  journal = {Proceedings of the Royal Society of London A},
  volume  = {356},
  pages   = {237--257},
  year    = {1977},
  doi     = {10.1098/rspa.1977.0130}
}

@article{HolzheyLarsenWilczek1994,
  author  = {Holzhey, Christoph and Larsen, Finn and Wilczek, Frank},
  title   = {Geometric and Renormalized Entropy in Conformal Field Theory},
  journal = {Nuclear Physics B},
  volume  = {424},
  pages   = {443--467},
  year    = {1994},
  doi     = {10.1016/0550-3213(94)90402-2},
  eprint  = {hep-th/9403108},
  archivePrefix = {arXiv}
}

@article{CalabreseCardy2004,
  author  = {Calabrese, Pasquale and Cardy, John},
  title   = {Entanglement Entropy and Quantum Field Theory},
  journal = {Journal of Statistical Mechanics: Theory and Experiment},
  volume  = {2004},
  pages   = {P06002},
  year    = {2004},
  doi     = {10.1088/1742-5468/2004/06/P06002},
  eprint  = {hep-th/0405152},
  archivePrefix = {arXiv}
}

@article{BianchiSmerlak2014,
  author  = {Bianchi, Eugenio and Smerlak, Matteo},
  title   = {Entanglement Entropy and Negative Energy in Two Dimensions},
  journal = {Physical Review D},
  volume  = {90},
  pages   = {041904},
  year    = {2014},
  doi     = {10.1103/PhysRevD.90.041904},
  eprint  = {1404.0602},
  archivePrefix = {arXiv},
  primaryClass = {hep-th}
}

@article{BianchiDeLorenzoSmerlak2015,
  author  = {Bianchi, Eugenio and De Lorenzo, Tommaso and Smerlak, Matteo},
  title   = {Entanglement Entropy Production in Gravitational Collapse: Covariant Regularization and Solvable Models},
  journal = {Journal of High Energy Physics},
  volume  = {06},
  pages   = {180},
  year    = {2015},
  doi     = {10.1007/JHEP06(2015)180},
  eprint  = {1409.0144},
  archivePrefix = {arXiv},
  primaryClass = {hep-th}
}

@article{AbdolrahimiPage2015,
  author  = {Abdolrahimi, Shohreh and Page, Don N.},
  title   = {Hawking Radiation Energy and Entropy from a Bianchi--Smerlak Semiclassical Black Hole},
  journal = {Physical Review D},
  volume  = {92},
  pages   = {083005},
  year    = {2015},
  doi     = {10.1103/PhysRevD.92.083005},
  eprint  = {1506.01018},
  archivePrefix = {arXiv},
  primaryClass = {hep-th}
}

@article{FordRoman1999,
  author  = {Ford, L. H. and Roman, Thomas A.},
  title   = {The Quantum Interest Conjecture},
  journal = {Physical Review D},
  volume  = {60},
  pages   = {104018},
  year    = {1999},
  doi     = {10.1103/PhysRevD.60.104018},
  eprint  = {gr-qc/9901074},
  archivePrefix = {arXiv}
}

@article{FewsterTeo2000,
  author  = {Fewster, Christopher J. and Teo, Edward},
  title   = {Quantum Inequalities and ``Quantum Interest'' as Eigenvalue Problems},
  journal = {Physical Review D},
  volume  = {61},
  pages   = {084012},
  year    = {2000},
  doi     = {10.1103/PhysRevD.61.084012},
  eprint  = {gr-qc/9908073},
  archivePrefix = {arXiv}
}

@article{FewsterHollands2005,
  author  = {Fewster, Christopher J. and Hollands, Stefan},
  title   = {Quantum Energy Inequalities in Two-Dimensional Conformal Field Theory},
  journal = {Reviews in Mathematical Physics},
  volume  = {17},
  pages   = {577--612},
  year    = {2005},
  doi     = {10.1142/S0129055X05002406},
  eprint  = {math-ph/0412028},
  archivePrefix = {arXiv}
}

@article{EckerEtAl2019,
  author  = {Ecker, Christian and Grumiller, Daniel and van der Schee, Wilke and Sheikh-Jabbari, Mohammad M. and Stanzer, Philipp},
  title   = {Quantum Null Energy Condition and its (Non)Saturation in 2D CFTs},
  journal = {SciPost Physics},
  volume  = {6},
  pages   = {036},
  year    = {2019},
  doi     = {10.21468/SciPostPhys.6.3.036},
  eprint  = {1901.04499},
  archivePrefix = {arXiv},
  primaryClass = {hep-th}
}

@article{BlancoEtAl2018,
  author  = {Blanco, David and Casini, Horacio and Leston, Mauricio and Rosso, Felipe},
  title   = {Modular Energy Inequalities from Relative Entropy},
  journal = {Journal of High Energy Physics},
  volume  = {01},
  pages   = {154},
  year    = {2018},
  doi     = {10.1007/JHEP01(2018)154},
  eprint  = {1711.04816},
  archivePrefix = {arXiv},
  primaryClass = {hep-th}
}

@article{KibeEtAl2022,
  author  = {Kibe, Tanay and Mukhopadhyay, Ayan and Roy, Pratik},
  title   = {Quantum Thermodynamics of Holographic Quenches and Bounds on the Growth of Entanglement from the Quantum Null Energy Condition},
  journal = {Physical Review Letters},
  volume  = {128},
  pages   = {191602},
  year    = {2022},
  doi     = {10.1103/PhysRevLett.128.191602},
  eprint  = {2109.09914},
  archivePrefix = {arXiv},
  primaryClass = {hep-th}
}

@article{KibeRoy2025,
  author  = {Kibe, Tanay and Roy, Pratik},
  title   = {Quantum Null Energy Condition in Quenched 2D CFTs},
  journal = {Physical Review D},
  volume  = {111},
  pages   = {126009},
  year    = {2025},
  doi     = {10.1103/2p7v-f28f},
  eprint  = {2503.17448},
  archivePrefix = {arXiv},
  primaryClass = {hep-th}
}

@misc{FlissRolph2025,
  author       = {Fliss, Jackson R. and Rolph, Andrew},
  title        = {Curious QNEIs from QNEC: New Bounds on Null Energy in Quantum Field Theory},
  year         = {2025},
  eprint       = {2510.26247},
  archivePrefix = {arXiv},
  primaryClass = {hep-th}
}

\end{document}